\documentclass[english]{article}
\usepackage[a4paper, total={6in, 8in}]{geometry}

\usepackage{amsmath}
\usepackage{amssymb}
\usepackage{amsthm}
\usepackage{mathtools}
\usepackage{enumitem}
\usepackage{relsize}

\usepackage{xcolor}
\usepackage[colorlinks=true, allcolors=blue!75!black]{hyperref}
\usepackage{tensor}

\usepackage[backend=biber, style=numeric-comp,sorting=none,giveninits=true]{biblatex} % set backend=bibtex, then biber
\AtEveryBibitem{\clearfield{isbn}} % remove isbn entries
\AtEveryBibitem{\clearfield{month}} % remove month entries

\numberwithin{equation}{section}
\newtheorem{theorem}{Theorem}
\newtheorem*{theorem*}{Theorem}

\newtheorem{proposition}{Proposition}
\newtheorem*{proposition*}{Proposition}

\newtheorem{definition}{Definition}
\newtheorem*{definition*}{Definition}

\usepackage[symbol]{footmisc}
\begin{document}

\begin{titlepage}
\thispagestyle{plain}
% to allow numbering from title page
\let\endtitlepage\relax
% to allow text after abstract

\begin{center}
    {\Large\bf Singularity and differentiability at the origin\\ of static and spherically symmetric black holes}\\\vspace{0.5cm}
    {\large Tommaso Antonelli\footnote[1]{\href{mailto:t.antonelli@sussex.ac.uk}{t.antonelli@sussex.ac.uk}} and Marco Sebastianutti\footnote[2]{\href{mailto:m.sebastianutti@sussex.ac.uk}{m.sebastianutti@sussex.ac.uk}} }\\\vspace{0.5cm}
    {\centering\it Department of Physics and Astronomy,\\ University of Sussex, Brighton, BN1 9QH, United Kingdom}\\\vspace{0.25cm}
\end{center}
\vspace{0.25cm}

\begin{abstract}
\noindent The divergence of curvature invariants at a given point signals the impossibility of extending the spacetime to that point, with the derivative order of these diverging invariants determining the differentiability class of the considered spacetime. 
We hereby focus on a general static and spherically symmetric geometry and determine, in the full non-linear regime and in a model-independent way, the conditions that the metric functions must satisfy in order to achieve finiteness of all curvature invariants at the origin. Our findings have direct implications regarding the extendibility of such spacetimes, which we illustrate by making explicit examples of various black hole geometries.
This work is structured around a central theorem, which relates the finiteness of curvature invariants at the origin to the leading order behavior and parity properties of the metric functions. The detailed proof of this theorem constitutes the main result of the paper.
\end{abstract}  
\end{titlepage}
\vspace{0.25cm}
\renewcommand{\thefootnote}{\arabic{footnote}}

\section{Introduction}
\label{sec:intro}
The singularity problem in general relativity is one of the most important theoretical challenges of the theory. Every classical black hole solution has a curvature singularity at its core~---~namely, a singularity signaled by the divergence of curvature tensors containing two derivatives of the metric, which cannot be removed by any change of coordinates. More in general, the singularity theorems~\cite{Penrose:1964wq,Hawking:1966sx,Hawking:1970zqf,Hawking:2023lss} (see~\cite{Senovilla:2021pdg} for a modern review) give sufficient conditions under which the spacetime can develop a geodesic singularity~---~that is, a singularity characterized by the existence of causal incomplete geodesics, in other words, timelike or null geodesics that cannot be extended to arbitrary values of their affine parameters.

Despite these theorems, the identification and physical interpretation of singularities in general relativity remains a subtle issue. On one hand, even when some incomplete geodesic curves are present, curvature may still remain everywhere finite. For instance, if one is solving the geodesic equation in a specific coordinate chart, the inextendibility of geodesics may be simply linked to the inextendibility of that coordinate patch, without any physical singularity, e.g.~geodesics in the Rindler patch of Minkowski space~\cite{Rindler:1960zz,Rindler:1966zz}; for a clear exposition of the geodesic incompleteness of Rindler spacetime, see~\cite{Wald:1984rg} Sec.~6.4. Moreover, these incomplete geodesic curves may correspond to a removable singularity, e.g.~Minkowski with a point removed. On the other hand, when curvature diverges in a region that can be reached in finite affine parameter (e.g.~Schwarzschild), we can conclude fairly generally that the spacetime is $C^k$-inextendible past a certain point~\cite{Ellis:1977pj}, where $k$ denotes the differentiability class of the metric. In this case, $k$ can be determined by: {\it i)} the derivative order of a diverging curvature invariant (see~\cite{Chrusciel:2020fql} Sec.~4.4.2); {\it ii)} the failure of a component of $\nabla_{\mu_1}\dots\nabla_{\mu_{k-2}}R_{\alpha\beta\gamma\delta}$ to remain finite upon evaluation in a parallelly propagated orthonormal (or null) tetrad along a causal geodesic~---~more generally, along a curve~---~approaching the singular point~\cite{Clarke:1973ex,Clarke:1982ex,Clarke:1982ld}, where $R_{\alpha\beta\gamma\delta}$ denotes the Riemann tensor. Note that {\it ii)} is a more general criterion than {\it i)} for detecting $C^k$-inextendibility. In fact, while {\it i)} implies {\it ii)}, the converse does not always hold.\footnote{To be more precise, {\it ii)} is satisfied if and only if the spacetime is $C^k$-inextendible past a certain point~\cite{Clarke:1973ex,Clarke:1982ex,Clarke:1982ld}. It is then possible to realize spacetimes in which curvature invariants with a given number of derivatives of the metric remain finite at the limiting point, but in which some parallelly propagated curvature components of the same derivative order diverge, see~\cite{Wald:1984rg} Sec.~9.1 and~\cite{Ellis:1977pj}.} For a more recent and exhaustive classification on the nature of spacetime singularities, see~\cite{Curiel:1999ss}. 

Even though geodesic singularities are more subtle in their physical interpretation than curvature singularities, the singularity theorems show that it is much more immediate to claim the geodesic incompleteness of a general spacetime rather than looking at its diverging curvature invariants.
In fact, without fixing any particular coordinate system, the singularity theorems state that geodesic singularities are implied by general assumptions, such as the presence of trapped surfaces and the requirement that an energy condition holds; but nothing is really said about curvature singularities.
However, in the present case, since we study static and spherically symmetric spacetimes, we can pick a convenient coordinate patch that considerably simplifies the analysis of the curvature scalars. We therefore focus solely on the finiteness (divergence) of these invariants, using them to draw conclusions about the $C^k$-(in)extendibility of the spacetime.

As mentioned above, all standard black hole solutions are singular at their core. Focusing on static and spherically symmetric spacetimes, Schwarzschild and Reissner--Nordstr\"om are two examples of singular geometries, both characterized by a diverging Krestchmann scalar $\tensor{R}{_{\mu\nu\rho\sigma}}\tensor{R}{^{\mu\nu\rho\sigma}}$ at $r=0$. In the spirit of producing a black hole solution that does not feature a singular core, beginning with the seminal~\cite{Bardeen:1968reg} and subsequent works~\cite{Poisson:1988wc,Frolov:1989pf,Dymnikova:1992ux,Mars:1996khm,Ayon-Beato:1998hmi,Ayon-Beato:2000mjt}, a variety of regular black hole models have been proposed in the literature~\cite{Bonanno:2000ep,Bronnikov:2000vy,Dymnikova:2001fb,Nicolini:2005vd,Hayward:2005gi,Bronnikov:2005gm,Lemos:2011dq,Bambi:2013ufa}. Recent developments, featuring contributions from various directions within gravitational physics, have stimulated renewed interest in this field. These include purely phenomenological models~\cite{Fan:2016hvf,Frolov:2017rjz,Simpson:2019mud}, models inspired by non-linear electrodynamics~\cite{Culetu:2013fsa,Culetu:2014lca,Balart:2014cga,DeFelice:2024seu}, models motivated by black-bounce mechanisms~\cite{Simpson:2018tsi,Lobo:2020ffi,Franzin:2021vnj,Bronnikov:2021uta,Bronnikov:2024izh}, those that implement quantum effects~\cite{DeLorenzo:2014pta,Nicolini:2019irw}, and several other proposals~\cite{Holdom:2002xy,Xiang:2013sza,Carballo-Rubio:2022kad,Calmet:2020vuh,Ovalle:2023ref,Ovalle:2024wtv,Casadio:2025pun,Boos:2024sgm} including the coherent state approach to quantum black holes~\cite{Casadio:2016zpl,Casadio:2017cdv,Casadio:2021eio,Casadio:2022ndh,Feng:2025nai,Antonelli:2025mcv}. For recent reviews on the subject, see~\cite{Bambi:2023try,Carballo-Rubio:2025fnc}.

The listed solutions are typically crafted to regularize second-derivative curvature invariants in general relativity, such as the Ricci scalar $R$ and Kretschmann scalar. Thus, despite being commonly referred to as ``regular black holes'' in the literature, as already pointed out in~\cite{Burzilla:2020utr,Giacchini:2021pmr}, these models may still yield divergences in higher-derivative curvature invariants, such as $\Box^N R$ and $\tensor{R}{_{\mu\nu\rho\sigma}}\Box^N\tensor{R}{^{\mu\nu\rho\sigma}}$. The importance of such invariants becomes apparent in the perturbative approach to quantum gravity where the action often includes these terms, originating, for example, from the renormalization of loop diagrams~\cite{Birrell:1982ix} or from the low-energy limits of theories that are expected to be valid at high energies, like string theory~\cite{Asorey:1996hz}. Clearly, these curvature invariants play a central role also in higher-derivative theories of gravity, including those with an infinite number of derivatives of the metric~\cite{Modesto:2011kw}. Furthermore, the appearance of higher-derivative curvature terms may be justified in light of the finite action principle~\cite{Barrow:1988act,Barrow:2019gzc,Lehners:2019ibe,Jonas:2021xkx}, since, as investigated in~\cite{Borissova:2020knn,Borissova:2023kzq,Borissova:2024hkc} for a static and spherically symmetric spacetime, when the action contains curvature tensors of higher order and the spacetime is $C^{k}$-inextendible to $r=0$ in the sense of {\it i)}, the divergence of these additional operators can lead to the divergence of the action functional, which, in turn, suppresses the contribution of this spacetime in the gravitational path integral.

In the present work, we analyze, in a model-independent way and in the full non-linear regime, the problem of the extendibility of a static and spherically symmetric spacetime to the point $r=0$. More concretely, we generalize and formalize the conjecture in~\cite{Giacchini:2021pmr} by putting it in the form of a theorem, to which we provide a rigorous proof. We show that all curvature invariants of arbitrarily high order are finite at $r=0$ if and only if the metric functions have specific leading order behaviors and parity properties at this point. Furthermore, building from the aforementioned, we elaborate on the $C^k$-extendibility of the spacetime to~$r=0$, where $k\in\mathbb{N}_0\cup\{\infty\}\cup\{\omega\}$, finding that the degree of differentiability at the origin of the considered geometry is dictated by the degree of ``evenness'' of the metric functions. Finally, we provide explicit examples illustrating how these results apply in the context of (regular) black hole spacetimes. At the same time, we emphasize the generality of our findings, which hold for a very broad class of static and spherically symmetric geometries.

The rest of this work is organized as follows: in Sec.~\ref{sec:thm} we introduce Thm.~\ref{thm:main}, the main theorem of the paper, which requires both directions of the ``if and only if'' claim to be proven. In Sec.~\ref{sec:rev} we prove the backward direction of the statement, and in Sec.~\ref{sec:dir} we prove the forward one. Then, in Sec.~\ref{sec:extensions} we generalize the results of the main theorem by presenting Thms.~\ref{thm:smooth},~\ref{thm:analytic} and~\ref{thm:Cn}. Finally, in Sec.~\ref{sec:conclusions} we provide some applications of the theorems and make our concluding remarks. In App.~\ref{sec:properties} we examine the properties of the class of functions that play a central role in the formulation of the theorems.

\section{Formulation of the theorem}\label{sec:thm}
We consider the general static and spherically symmetric geometry described by the metric
\begin{equation}\label{eq:metric}
  ds^2=-A(r)\,dt^2+B(r)\,dr^2+r^2\,d\Omega^2,
\end{equation}
where $d\Omega^2$ is the line element on the unit two-sphere and $A(r)$ and $B(r)$ are functions of the areal radius~$r$. Two remarks regarding the above metric are in order, see e.g.~Secs.~10.1 and~10.2 of~\cite{Schutz:2022gr}:

{\it i)} For Eq.~\eqref{eq:metric} to be static, it is required that $A(r)>0$; if not, $\partial_t$, the Killing vector that generates time translations, would not be timelike. Given the fact that black hole metrics have spacetime regions where $A(r)<0$~---~e.g.~the region inside the event horizon of a Schwarzschild black hole~---~we do not impose $A(r)$ to be positive. Nevertheless, we improperly still refer to these geometries as ``static'', with this term only denoting the invariance of the metric under translations and reflections of the coordinate $t$.

{\it ii)} Furthermore, while it is true that every static and spherically symmetric metric can be cast in the above form~---~e.g.~the angular part $C(r')\,d\Omega^2$ can always be brought into the form of Eq.~\eqref{eq:metric} by redefining the radial coordinate as $r(r')\equiv\sqrt{C(r')}$~---~it may happen that the areal radius' domain is restricted to $r\ge r_0>0$, as in the wormhole geometry of~\cite{Morris:1988cz} wherein $C(r')=r'^2+r_0^2$, or that the point $r=0$ cannot be reached in a finite amount of affine parameter~\cite{Bambi:2016wdn}. In what follows, we exclude such scenarios and consider only static and spherically symmetric spacetimes in which the point $r=0$ is accessible.\bigbreak

For the purposes of this work, {\it we restrict $A(r)$ and $B(r)$ to be of the form
\begin{equation}\label{eq:ansatz_metric_functions}
    A(r)=r^m \tilde{A}(r), \quad B(r)=r^n \tilde{B}(r), \quad m,n\in\mathbb{R},\quad \tilde{A}(0)\neq0,\quad \tilde{B}(0)\neq0,
\end{equation}
where $\tilde{A}(r)$ and $\tilde{B}(r)$ are smooth functions in a neighborhood of $r=0$}. Additionally, we adopt the following notation for the derivatives of $\tilde{A}(r)$ and $\tilde{B}(r)$ at the origin:
\begin{equation}
  a_k\equiv\frac{\tilde{A}^{(k)}(0)}{k!},\quad b_k\equiv\frac{\tilde{B}^{(k)}(0)}{k!}, \quad \forall k\in\mathbb{N}_0.
\end{equation}
Note that the parameters $m$ and $n$ represent the leading order behaviors at $r=0$ of $A(r)$ and $B(r)$, respectively, hence the assumptions $a_0\neq0$ and $b_0\neq0$.

In the context we are interested in, the assumptions imposed above on $A(r)$ and $B(r)$ are minimal: we are mainly excluding $\tilde{A}(r)$, $\tilde{B}(r)$ that are not smooth around $r=0$ and $A(r)$, $B(r)$ that are rapidly increasing (decreasing) functions at the origin. By the latter terminology, we refer to functions $f(x)$ that tend to infinity (zero) faster than any power $x^\alpha$ for any $\alpha\in\mathbb{R}$ with $\alpha<0$ ($\alpha>0$) as $x\to0$. Analogously, by non-rapidly increasing (decreasing) $f(x)$, we refer to functions that behave as ${\cal O}\!\left(x^{\alpha}\right)$, where again $\alpha\in\mathbb{R}$ and $\alpha<0$ ($\alpha>0$) as $x\to0$.

As stated in Eq.~\eqref{eq:ansatz_metric_functions}, we are considering functions $A(r)$ that can be represented around $r=0$ as the product of $r^m$ and a smooth non-vanishing function $\tilde{A}(r)$; analogously for $B(r)$. These $A(r)$, in the limit $r\to0$, can either tend to zero (if $m>0$), infinity (if $m<0$), or a non-zero value (if $m=0$). Note that functions $\tilde{A}(r)$ like $\exp(\pm1/r^2)$, which as $r\to0$ tend to infinity or zero faster than any polynomial, are excluded by our assumptions; the same goes for $\tilde{A}(r)=r^{\alpha}\exp(\pm1/r^2)$, $\forall\alpha\in\mathbb{R}$. However, functions $\tilde{A}(r)$ such as $1+\exp(-1/r^2)$, which are smooth (but non-analytic) around $r=0$ and have a non-zero limit at that point, constitute valid $\tilde{A}(r)$.\footnote{The theorems we present in this paper are valid only for metric functions that satisfy these conditions, although we suspect our results to hold for rapidly increasing and decreasing functions as well. In the latter case, since we cannot use Taylor's theorem to approximate the metric functions as non-vanishing polynomials around the origin, a much more technical treatment from the one presented here might be required.}\bigbreak

In the following discussion, we work with smooth functions with specific parity properties at the origin. In the context of smooth functions, it turns out that the usual notions of even and odd functions are too restrictive for our purposes; therefore, we introduce the concepts of d-evenness and d-oddness (where ``d'' stands for derivative):
\begin{definition}\label{def:d-parity}
Given a function $f(x)$ that is smooth in a neighborhood of $x=0$, we say $f(x)$ is d-even if~$f^{(2k+1)}(0)=0$, similarly we say $f(x)$ is d-odd if $f^{(2k)}(0)=0,\ \forall k\in\mathbb{N}_0$.
\end{definition}
\noindent For smooth functions, if $f(x)$ is even (odd) then it is d-even (d-odd), the converse being, in general, not true. For analytic functions, instead, the concepts of d-evenness and d-oddness coincide with the standard notions of evenness and oddness.\bigbreak

We can now formalize the conjecture in~\cite{Giacchini:2021pmr} and present it as a theorem:
\begin{theorem}
  \label{thm:main}
  Given the metric in Eq.~\eqref{eq:metric}, where $A(r)$ and $B(r)$ satisfy the assumptions in Eq.~\eqref{eq:ansatz_metric_functions}, all curvature invariants are finite at $r=0$ if and only if $m=n=0$, $b_0=1$, and $\tilde{A}(r)$, $\tilde{B}(r)$ are d-even functions of $r$.
\end{theorem}

\noindent We specify that, by ``all curvature invariants'' we mean any scalar built from a contraction of an arbitrary number of curvature tensors and covariant derivatives of the said tensors, such as $\Box^N R$, $\nabla_{\mu_1}\dots\nabla_{\mu_N}\tensor{R}{^\rho_\sigma}\nabla^{\mu_1}\dots\nabla^{\mu_N}\tensor{R}{^\sigma_\rho}$, $\tensor{R}{_{\mu\nu\rho\sigma}}\Box^N\tensor{R}{^{\mu\nu\rho\sigma}}$, and so on; including curvature invariants obtained from contractions with the Levi-Civita tensor $\tensor{\varepsilon}{_{\mu\nu\rho\sigma}}$.

As anticipated in Sec.~\ref{sec:intro}, both directions of the statement must be proved in order to prove the theorem. In Sec.~\ref{sec:rev} we prove the backward direction, i.e.~if ``$m=n=0$, $b_0=1$, and $\tilde{A}(r)$, $\tilde{B}(r)$ are d-even functions of $r$'' then ``all curvature invariants are finite at $r=0$''. In Sec.~\ref{sec:dir} we prove the forward direction, i.e.~if ``all curvature invariants are finite at $r=0$'' then ``$m=n=0$, $b_0=1$, and $\tilde{A}(r)$, $\tilde{B}(r)$ are d-even functions of $r$''. Essential for the proof of the latter direction are the Taylor expansions, around $r=0$, of several higher-derivative curvature invariants. These have been computed in \textsc{Mathematica} using the \texttt{xAct} package~\cite{Martin-Garcia:2025act}.

\section{Proof of the backward direction}\label{sec:rev}
Under the assumptions of the theorem, we prove that if $m=n=0$, $b_0=1$, and $\tilde{A}(r)$, $\tilde{B}(r)$ are d-even functions, then we can construct a smooth coordinate chart in a neighborhood of $r=0$, which, in turn, implies that all curvature invariants are finite at that point. 

In fact, all curvature invariants can be regarded as derivatives of some order of the metric contracted with the inverse metric. Thus, if we are able to find a smooth coordinate chart around the point that corresponds, in spherical coordinates, to $r=0$, then all derivatives of the metric of arbitrarily high order are well defined and finite at this point, as are those of the inverse metric. It follows that any possible combination of the two is also finite there, implying that every curvature scalar, whose value at a point is coordinate independent, remains finite in any coordinate system.\bigbreak

Given the form of the metric in Eq.~\eqref{eq:metric}, we can define a new radius $\tilde{r}$ to satisfy the following differential equation (isotropic coordinates):
\begin{equation}\label{eq:change_radial}
  B(r)\biggl(\frac{dr}{d\tilde{r}}\biggr)^2=
  \frac{r^2}{\tilde{r}^2}\equiv C(r).
\end{equation}
If this coordinate transformation is well defined, it has the effect of turning the metric into
\begin{equation}
  ds^2=-A(r)\,dt^2+C(r)\!\left(d\tilde{r}^2+\tilde{r}^2d\Omega^2\right)\!,
\end{equation}
which, in turn, after the change of coordinates
\begin{equation}\label{eq:polar_to_cartesian}
    \begin{dcases}
        x= \tilde{r} \sin\theta\cos\phi\\
        y= \tilde{r} \sin\theta\sin\phi\\
        z=\tilde{r}\cos\theta,
    \end{dcases}
\end{equation}
becomes simply
\begin{equation}
  \label{eq:new_metric}
  ds^2=-A(r)\,dt^2+C(r)\!\left(dx^2+dy^2+dz^2\right)\!.
\end{equation}
The Cartesian coordinates have the effect of removing the $\tilde{r}^2$ factor from the metric, which is a source of ill definition of the spherical coordinates at the origin. Since $m=0$, $A(r)\equiv\tilde{A}(r)$ with $A(0)\equiv a_0\neq0$. Hence, the metric is non-degenerate at $r=0$ if and only if $C(0)$ is finite and non-zero.

We now prove that, under the conditions of Thm.~\ref{thm:main}, this change of coordinates is non-singular and produces a smooth metric around $r=0$. To this end, we first analyze when the change of radial coordinate from $r$ to $\tilde{r}$ yields a non-singular metric at $r=0$.

From Eq.~\eqref{eq:change_radial} we obtain
\begin{equation}
  \label{eq:diff_eq_for_radial}
  \frac{d\tilde{r}}{\tilde{r}}=\frac{dr}{r}\sqrt{B(r)},
\end{equation}
which can be integrated to produce
\begin{equation}\label{eq:}
  \tilde{r}(r) =\exp\!\left( \int\frac{dr}{r}\sqrt{B(r)}\right)\!,
\end{equation}
where, since $n=0$, $B(r)\equiv \tilde{B}(r)$ with $B(0)\equiv b_0\neq0$. By Taylor expanding $B(r)$ around $r=0$ and choosing the constant of integration to be zero in the above, $\tilde{r}(r)$ can be expanded in a neighborhood of $r=0$ as
\begin{equation}
  \label{eq:rtilde}
  \tilde{r}(r)=r^{\sqrt{b_0}}\!\left(1+{\cal O}(r)\right)\!.
\end{equation}
Inserting this expression into $C(r)$, we find
\begin{equation}
  C(r)=\frac{r^2}{\tilde{r}(r)^2}=r^{2(1-\sqrt{b_0})}(1+{\cal O}(r)),
\end{equation}
and taking the limit $r\to0$, we obtain
\begin{equation}
  \lim_{r\to0}C(r)=1\neq 0\iff b_0=1,
\end{equation}
which shows how the assumption $b_0=1$ makes the metric in Eq.~\eqref{eq:new_metric} non-degenerate at $r=0$. From now on, we enforce this condition.

We can also see, from Eq.~\eqref{eq:change_radial} and since $C(0)\neq0$, that the limit $r\to 0$ corresponds to $\tilde{r}\to 0$, in turn, corresponding to $(x,y,z)\to (0,0,0)$ given Eqs.~\eqref{eq:polar_to_cartesian}, which makes $(0,0,0)$ a perfectly valid point in the chosen coordinate chart.

We have thus obtained a non-singular coordinate neighborhood of $r=0$. However, we still have to prove that this coordinate neighborhood is smooth around $r=0$, which is not clearly apparent from Eq.~\eqref{eq:new_metric}, given that the dependence of $A$ and $C$ on $(x,y,z)$ is still rather implicit. To this end, we study the smoothness and parity properties of these functions.\bigbreak

With the choice of $b_0=1$, $\tilde{r}(r)$ depends smoothly on $r$; and since $\tilde{r}'(0)=1\neq0$, by the inverse function theorem also the inverse function $r(\tilde{r})$ is smooth at $\tilde{r}=0$, see e.g.~\cite{Tao:2022ana} Sec.~6.7 and~\cite{Hormander:2003pdo} Thm.~1.1.7. Moreover, from Eq.~\eqref{eq:change_radial}, it is immediate to check that $C(r)$ depends smoothly on $r$ and, since $r(\tilde{r})$ is a smooth function, $C$ depends smoothly also on $\tilde{r}$.

To determine the parity of the functions here involved, it is important to check that the properties of d-even and d-odd functions, see Def.~\ref{def:d-parity}, coincide with those of standard even and odd functions. The results can be summarized in the following proposition (which we prove in App.~\ref{sec:properties}):
\begin{proposition}\label{prop:d-parity}
Given $f$ and $g$ smooth functions, if necessary equipped with a smooth inverse:
\begin{enumerate}[label=\alph*)]
  \item $f$ generic, $g$ d-even $\ \implies\ $ $f\circ g$ d-even;
  \item $f$ d-even, $g$ d-odd $\ \implies\ $ $f\circ g$ d-even;
  \item $f$ d-odd, $g$ d-odd $\ \implies\ $ $f\circ g$ d-odd;
  \item $f$ d-odd $\ \iff\ $ $f^{-1}$ d-odd;
  \item $f$ d-even $\ \iff\ $ $f'$ d-odd;
  \item $f$ d-odd $\ \implies\ $ $f'$ d-even.
\end{enumerate}
\end{proposition}

\noindent Since $B(r)$ is assumed to be a smooth and d-even function of $r$, using the above properties, it is possible to see that $\tilde{r}(r)$ is smooth and d-odd, which, in turn, implies that $C(r)$ is smooth and d-even. In summary, both $A(r)$ and $C(r)$ are smooth d-even functions of $r$.

Since $\tilde{r}(r)$ is a smooth d-odd function of $r$, by Prop.~\ref{prop:d-parity} we can immediately infer that also the inverse function $r(\tilde{r})$ is a smooth d-odd function of $\tilde{r}$. As a consequence of that, again using the above properties, the compositions $A(r(\tilde{r}))$ and $C(r(\tilde{r}))$ are smooth d-even functions of $\tilde{r}$ as well.

This means that the Taylor expansions of  $A(r(\tilde{r}))$ and $C(r(\tilde{r}))$ involve only even powers of $\tilde{r}$, and given the relation
\begin{equation}
  \label{eq:xyz}
  \tilde{r}^2=x^2+y^2+z^2,
\end{equation}
it follows that the Taylor expansions of $A(r(\tilde{r}(x,y,z)))$ and $C(r(\tilde{r}(x,y,z)))$ are polynomials in $x^2$, $y^2$ and $z^2$. In turn, all partial derivatives of these functions w.r.t.~$x$, $y$ and $z$ are well defined at $\tilde{r}=0$, meaning that $A(r(\tilde{r}(x,y,z)))$ and $C(r(\tilde{r}(x,y,z)))$ are smooth (and d-even) functions also in this coordinate chart. Note that the smoothness would not be achieved if there were odd powers of $\tilde{r}$ in the Taylor expansions of $A(r(\tilde{r}))$ and $C(r(\tilde{r}))$, since $\tilde{r}$ itself does not depend smoothly on $(x,y,z)$ at $\tilde{r}=0$.

The above discussion implies that the metric in Eq.~\eqref{eq:new_metric} can be written as
\begin{equation}\label{eq:extension}
    ds^2=-A(r(\tilde{r}(x,y,z)))\,dt^2+C(r(\tilde{r}(x,y,z)))\!\left(dx^2+dy^2+dz^2\right)\!,
\end{equation}
where the dependence of $A$ and $C$ on $(x,y,z)$ is smooth, and the metric is now clearly smooth around $r=0$. This proves our claim.\bigbreak 

We emphasize that, given the metric in Eq.~\eqref{eq:metric}, the one in Eq.~\eqref{eq:extension} provides an explicit $C^\infty$-extension of the spacetime to $r=0$. More generally, by slightly modifying the assumptions on the metric functions $A(r)$ and $B(r)$, we later show (see Thms.~\ref{thm:smooth},~\ref{thm:analytic} and~\ref{thm:Cn}) that the same analysis presented in this section yields different degrees of $C^k$-extendibility to $r=0$, where $k\in\mathbb{N}_0\cup\{\infty\}\cup\{\omega\}$.

\section{Proof of the forward direction}\label{sec:dir}
Under the assumptions of the theorem we show that if all curvature invariants are finite at the origin, then we find $m=n=0$, $b_0=1$, and $\tilde{A}(r)$, $\tilde{B}(r)$ d-even functions. Following the approach of~\cite{Giacchini:2021pmr}, we do this by proving the contrapositive statement. Namely, we demonstrate that if any of the conditions $m=n=0$, $b_0=1$, and $\tilde{A}(r)$, $\tilde{B}(r)$ d-even functions is not met, then we can always find a curvature invariant that diverges at $r=0$.\bigbreak

We begin our analysis by proving the necessity of the conditions $m=n=0$ and $b_0=1$. Under the assumptions of Eq.~\eqref{eq:ansatz_metric_functions}, if we suppose, by way of contradiction, that $n<0$, then the Ricci scalar expanded at $r=0$ reads
\begin{equation}
  R=2\,r^{-2}+{\cal O}\!\left(r^{-1}\right)\!.
\end{equation}
Therefore, since $n<0$ leads to a divergent curvature invariant, it follows that one must impose $n\geq0$.

If instead we suppose that $n>0$, then the expansions of the Ricci and Kretschmann scalars around $r=0$ read, respectively:
\begin{equation}
  \label{eq:ricci_alpha_beta} R=\frac{mn+4n-m^2-2m-4}{2b_0}\,r^{-2-n}+{\cal O}\!\left(r^{-1-n}\right)\!,
\end{equation}
\begin{equation}
  \label{eq:kretsch_alpha_beta} R_{\mu\nu\rho\sigma}R^{\mu\nu\rho\sigma}=\frac{m^2n^2+8n^2-2m^3n+4m^2n+m^4-4m^3+12m^2+16}{4b_0^2}\,r^{-4-2n}+{\cal O}\!\left(r^{-3-2n}\right)\!.
\end{equation}
In order to avoid the divergence in Eq.~\eqref{eq:ricci_alpha_beta}, we have to impose
\begin{equation}
  n=\frac{m^2+2m+4}{m+4},
\end{equation}
and, if we substitute the above in the coefficient of the diverging term in Eq.~\eqref{eq:kretsch_alpha_beta}, to set this coefficient to zero as well, we get the following quartic equation to solve:
\begin{equation}
  m^4+6m^3+24m^2+16m+24=0,
\end{equation}
for which no real solution exists. This means that our starting assumption $n>0$ also leads to a diverging curvature invariant. Therefore, from now on, we impose $n=0$.

If we now expand the Ricci and Kretschmann scalars under the assumption that $n=0$,  we obtain, respectively:
\begin{equation}
  \label{eq:ricci_alpha_b0} R=\frac{4b_0-m^2-2m-4}{2b_0}\,r^{-2}+{\cal O}\!\left(r^{-1}\right)\!,
\end{equation}
\begin{equation}
  \label{eq:kretsch_alpha_b0} R_{\mu\nu\rho\sigma}R^{\mu\nu\rho\sigma}=\frac{16b_0^2-32b_0+m^4-4m^3+12m^2+16}{4b_0^2}\,r^{-4}+{\cal O}\!\left(r^{-3}\right)\!.
\end{equation}
In order to remove the divergent term in Eq.~\eqref{eq:ricci_alpha_b0}, we have to impose
\begin{equation}
  \label{eq:b0_alpha}b_0=\frac{m^2+2m+4}{4},
\end{equation}
and again, if we substitute the above in Eq.~\eqref{eq:kretsch_alpha_b0}, to get rid of the divergent term, we have to solve the following quartic equation:
\begin{equation}
  m^2(m^2+8)=0,
\end{equation}
which has the only real solution $m=0$.

We can thus conclude that, in order to avoid the above divergences at $r=0$ in the Ricci and Kretschmann scalars, it is necessary to impose $m=n=0$, hence $A(r)\equiv\tilde{A}(r)$ and $B(r)\equiv\tilde{B}(r)$. Moreover, if we look at Eq.~\eqref{eq:b0_alpha}, we can see that $m=0$ implies another condition that must hold in order to ensure the finiteness at the origin of both $R$ and $R_{\mu\nu\rho\sigma}R^{\mu\nu\rho\sigma}$, namely, $b_0=1$. Therefore, for the rest of the discussion, we have that $A(r)$ and $B(r)$ are smooth functions with $a_0\neq0$ and $b_0=1$. Note that $a_0$ is not really a free parameter since, by the rescaling $t\mapsto|a_0|^{-1/2}\,t$ in Eq.~\eqref{eq:metric}, we can always set $a_0=\pm1$ without loss of generality. In most cases of physical interest, one typically fixes $a_0=1$ to ensure that the spacetime has a Lorentzian signature. However, since the following computations remain unchanged even when $a_0=-1$ (Euclidean signature), for the sake of generality, we keep the sign of $a_0$ unspecified.
\bigbreak

We continue our analysis by proving now the necessity of having $A(r)$, $B(r)$ d-even functions. Suppose, again by way of contradiction, to have some odd power in the Taylor expansions of $A(r)$ and $B(r)$ with a non-vanishing coefficient, i.e.
\begin{equation}\label{eq:expansion_AB}
  A(r)\sim a_{2n_a+1} \,r^{2n_a+1}, \quad B(r)\sim b_{2n_b+1}\,r^{2n_b+1},\quad N\equiv\min\{n_a,n_b\},
\end{equation}
where, in this notation, ``$\sim$'' denotes the first non-vanishing odd order term in the Taylor expansion of a quantity at $r=0$, and $n_a,n_b\in\mathbb{N}_0$.

From the analysis presented in~\cite{Giacchini:2021pmr}, the first odd order term in the Ricci scalar is
\begin{equation}
    R\sim-2(N+1)\!\left[(2N+1)\,\frac{a_{2N+1}}{a_0}-2b_{2N+1}\right]r^{2N-1},
\end{equation}
where $a_{2N+1}=0$ if $n_a>n_b$, and $b_{2N+1}=0$ if $n_a<n_b$. Acting on $R$ with the operator $\Box^N$, we obtain
\begin{equation}
  \Box^NR=-2(N+1)(2N)!\!\left[(2N+1)\,\frac{a_{2N+1}}{a_0}-2b_{2N+1}\right]r^{-1}+{\cal O}(1).
\end{equation}
Therefore, $\Box^N \tensor{R}{}$ diverges as $r^{-1}$ at $r=0$ unless 
\begin{equation}
  \label{eq:relation}
    a_{2N+1}=\frac{2a_0}{2N+1}b_{2N+1}.
\end{equation}
Note that, in the case where $n_a\neq n_b$, one and only one of $a_{2N+1}$ and $b_{2N+1}$ vanishes. In this scenario, Eq.~\eqref{eq:relation} does not apply and $\Box^N R$ is always divergent. 
Thus, if $n_a\neq n_b$, the proof of this direction of the statement terminates here. However, if $n_a=n_b=N$, the relation in Eq.~\eqref{eq:relation} may be satisfied, in which case we must continue our discussion. In the following, we impose the condition $n_a=n_b=N$.

Consequently, given the above result, our goal is to find another curvature invariant that diverges at $r=0$ unless a certain relation, independent from Eq.~\eqref{eq:relation}, between $a_{2N+1}$ and $b_{2N+1}$ is met. 
Should that relation be found, no matter what the relation between $a_{2N+1}$ and $b_{2N+1}$ is, if at least one of the coefficients does not vanish, then we would have proven that we can produce a diverging curvature invariant.

Let us start by exploring the Ricci tensor squared, which, at $r=0$ and for $N=0$, goes like
\begin{equation}
    \tensor{R}{^\mu_\nu}\tensor{R}{^\nu_\mu}=\!\left(\frac{3}{2}\frac{a_1^2}{a_0^2}-3\frac{a_1}{a_0}b_1+\frac{11}{2}b_1^2\right)r^{-2}+{\cal O}\!\left(r^{-1} \right)\!,
\end{equation}
after inserting~\eqref{eq:relation}, we still obtain a diverging curvature invariant
\begin{equation}
    \tensor{R}{^\mu_\nu}\tensor{R}{^\nu_\mu}=\frac{11}{2}b_1^2\,r^{-2}+{\cal O}\!\left(r^{-1} \right)\!.
\end{equation}
This tells us that if at least one of $a_1$ and $b_1$ is non-vanishing, no matter the relation between them, we obtain a diverging curvature invariant. For a generic $N>0$, the situation is a little more delicate, since the first odd order term in $\tensor{R}{^\mu_\nu}\tensor{R}{^\nu_\mu}$ is
\begin{equation}
    \tensor{R}{^\mu_\nu}\tensor{R}{^\nu_\mu}\sim\!\left[4(N+1)(2N+1)\!\left(2\frac{a_2}{a_0}-b_2\right)\!\frac{a_{2N+1}}{a_0}-8(N+1)\!\left(\frac{a_2}{a_0}-2b_2\right)\!b_{2N+1}\right]r^{2N-1},
\end{equation}
which, upon the insertion of~\eqref{eq:relation}, becomes
\begin{equation}
    \tensor{R}{^\mu_\nu}\tensor{R}{^\nu_\mu}\sim8(N+1)\!\left(\frac{a_2}{a_0}+b_2\right)\!b_{2N+1}\,r^{2N-1}.
\end{equation}
Unlike for $\Box^N R$, the $r^{-1}$ singularity in $\Box^N (\tensor{R}{^\mu_\nu}\tensor{R}{^\nu_\mu})$ can also be canceled by setting the even coefficients to $a_2=-a_0b_2$, which is not ideal in order to achieve a relation solely between $a_{2N+1}$ and $b_{2N+1}$. The same goes for $\Box^N (R_{\mu\nu\rho\sigma}R^{\mu\nu\rho\sigma})$ and for $\Box^N$ acting on other scalar quantities. Motivated by this, we go on exploring other types of tensor contractions. 

A curvature invariant that diverges at $r=0$ and produces a relation solely between $a_{2N+1}$ and $b_{2N+1}$, independent from Eq.~\eqref{eq:relation}, is $\Box^N\tensor{R}{^\mu_\nu}\,\Box^N\tensor{R}{^\nu_\mu}$. Observe that, for the metric in Eq.~\eqref{eq:metric}, the Ricci tensor $\tensor{R}{^\mu_\nu}$ is diagonal, with entries depending only on $r$:
\begin{align}\label{eq:ricci_components}
  \tensor{R}{^t_t}&=\frac{1}{B}\!\left(-\frac{A''}{2A}+\frac{A'^{\hspace{0.5mm}2}}{4A^2}+\frac{A'B'}{4AB}-\frac{A'}{Ar}\right)\nonumber\\
  \tensor{R}{^r_r}&=\frac{1}{B}\!\left(-\frac{A''}{2A}+\frac{A'^{\hspace{0.5mm}2}}{4A^2}+\frac{A'B'}{4AB}+\frac{B'}{Br}\right)\\
  \tensor{R}{^\theta_\theta}&=\tensor{R}{^\phi_\phi}=\frac{1}{B}\!\left(\frac{B-1}{r^2}-\frac{A'}{2Ar}+\frac{B'}{2Br}\right)\!,\nonumber
\end{align}
and with the first odd order terms in the Taylor expansions of its components at $r=0$ that read
\begin{align}\label{eq:ricci_components_odd}
    \tensor{R}{^t_t}&\sim-(2N+1)(N+1)\,\frac{a_{2N+1}}{a_0}\,r^{2N-1}\nonumber\\
    \tensor{R}{^r_r}&\sim(2N+1)\!\left(-N\frac{a_{2N+1}}{a_0}+b_{2N+1}\right)r^{2N-1}\\
    \tensor{R}{^\theta_\theta}&\sim\frac{1}{2}\!\left(-(2N+1)\frac{a_{2N+1}}{a_0}+(2N+3)b_{2N+1}\right)r^{2N-1}.\nonumber
\end{align}
Now, motivated by our interest in the curvature invariant $\Box^N\tensor{R}{^\mu_\nu}\,\Box^N\tensor{R}{^\nu_\mu}$, we study the way in which $\Box$ acts on a rank-two tensor $\tensor{F}{^\mu_\nu}$ that is of the same form as the Ricci tensor. To this end, if we take
\begin{equation}
  \label{eq:tensor_structure}
  \tensor{F}{^\mu_\nu}=\text{diag}\bigl(f_0(r),f_1(r),f_2(r),f_2(r)\bigr),
\end{equation}
we find that $\Box$ preserves the structure of the tensor, i.e. 
\begin{equation}
  \Box\tensor{F}{^\mu_\nu}=\text{diag}\bigl(\bar{f}_0(r),\bar{f}_1(r),\bar{f}_2(r),\bar{f}_2(r)\bigr),
\end{equation}
where
\begin{align}
  \label{eq:f0}
  &\bar{f}_0=\frac{1}{B}\!\left(f_0''+\frac{2f_0'}{r}+\!\left(\frac{A'}{2A}-\frac{B'}{2B}\right)\!f_0'-\frac{A'^{\hspace{0.5mm}2}}{2A^2}(f_0-f_1)\right)\!,\\
  \label{eq:f1}
  &\bar{f}_1=\frac{1}{B}\!\left(f_1''+\frac{2f_1'}{r}+\!\left(\frac{A'}{2A}-\frac{B'}{2B}\right)\!f_1'-\frac{A'^{\hspace{0.5mm}2}}{2A^2}(f_1-f_0)-\frac{4}{r^2}(f_1-f_2)\right)\!,\\
  \label{eq:f2}
  &\bar{f}_2=\frac{1}{B}\!\left(f_2''+\frac{2f_2'}{r}+\!\left(\frac{A'}{2A}-\frac{B'}{2B}\right)\!f_2'-\frac{2}{r^2}(f_2-f_1)\right)\!.
\end{align}
Therefore, if the expansions of the components of $\tensor{R}{^\mu_\nu}$ have their first odd power of~$r$ at order $2N-1$, it follows that the components of $\Box^N\tensor{R}{^\mu_\nu}$ diverge as $r^{-1}$, implying that the scalar $\Box^N\tensor{R}{^\mu_\nu}\,\Box^N\tensor{R}{^\nu_\mu}$ diverges as $r^{-2}$ at $r=0$. The only thing we are required to ensure is that the coefficient in front of the diverging term does not vanish, even when Eq.~\eqref{eq:relation} holds.

To check this, we need to study the coefficients of the first odd order terms in the expansions of the components of $\Box^{N-k} \tensor{R}{^\mu_\nu}$, where $k\in\{0,1,\dots N\}$. Since we know that every tensor of the type $\Box^{N-k} \tensor{R}{^\mu_\nu}$ has the same structure as in Eq.~\eqref{eq:tensor_structure}, we define $\alpha_k$, $\beta_k$ and $\gamma_k$ as the coefficients of the first odd powers in the components of this tensor:
\begin{equation}
  \label{eq:odd_term_box_ricci}
  \Box^{N-k} \tensor{R}{^\mu_\nu}\sim\text{diag}(\alpha_k,\beta_k,\gamma_k,\gamma_k)\,r^{2k-1}.
\end{equation}
Now, from the components of the Ricci tensor in Eqs.~\eqref{eq:ricci_components_odd}, we have that 
\begin{align}\label{eq:initialCs}
    \alpha_N&=-(2N+1)(N+1)\,\frac{a_{2N+1}}{a_0}\nonumber\\
    \beta_N&=(2N+1)\!\left(-N\frac{a_{2N+1}}{a_0}+b_{2N+1}\right)\\
    \gamma_N&=\frac{1}{2}\!\left(-(2N+1)\frac{a_{2N+1}}{a_0}+(2N+3)b_{2N+1}\right)\!,\nonumber
\end{align}
and from Eqs.~\eqref{eq:f0},~\eqref{eq:f1} and~\eqref{eq:f2}, we get the following recurrence relations:
\begin{equation}
  \begin{pmatrix}
    \alpha_{k-1}\\ \beta_{k-1}\\ \gamma_{k-1}
  \end{pmatrix} =
  M_{k}
  \begin{pmatrix}
    \alpha_{k}\\ \beta_{k}\\ \gamma_{k}
  \end{pmatrix},
\end{equation}
where $M_{k}$ denotes the matrix
\begin{equation}
  M_{k}=
  \begin{pmatrix}
    2k(2k-1) & 0 & 0 \\
    0 & 2(2k^2-k-2) & 4\\
    0 & 2 & 2(2k^2-k-1)
  \end{pmatrix},
\end{equation}
which can be diagonalized as $M_{k}=PD_{k}P^{-1}$, with the matrices
\begin{equation}\label{eq:matrixPDk}
  P=
  \begin{pmatrix}
    1 & 0 & 0\\
    0 & 1 & -2\\
    0 & 1 & 1
  \end{pmatrix}, \quad
  D_{k}=
  \begin{pmatrix}
    2k(2k-1) & 0 & 0\\
    0 & 2k(2k-1) & 0\\
    0 & 0 & 2(k+1)(2k-3)
  \end{pmatrix}.
\end{equation}
So, if we define
\begin{equation}
  \label{eq:relationCs}
  \begin{pmatrix}
    \tilde{\alpha}_{k}\\ \tilde{\beta}_{k}\\ \tilde{\gamma}_{k}
  \end{pmatrix}
  = P^{-1} 
  \begin{pmatrix}
    \alpha_{k}\\ \beta_{k}\\ \gamma_{k}
  \end{pmatrix}
  =
  \begin{pmatrix}
    1 & 0 & 0\\[2pt]
    0 & \dfrac{1}{3} & \dfrac{2}{3}\\[8pt]
    0 & -\dfrac{1}{3} & \dfrac{1}{3}
  \end{pmatrix}
  \begin{pmatrix}
    \alpha_{k}\\ \beta_{k}\\ \gamma_{k}
  \end{pmatrix},
\end{equation}
we obtain a simpler recurrence relation for the new coefficients $\tilde{\alpha}_k$, $\tilde{\beta}_k$ and $\tilde{\gamma}_k$, given by the diagonal matrix $D_{k}$:
\begin{equation}
  \label{eq:new_recurrence}
  \begin{pmatrix}
    \tilde{\alpha}_{k-1}\\ \tilde{\beta}_{k-1}\\ \tilde{\gamma}_{k-1}
  \end{pmatrix} =
  D_{k}
  \begin{pmatrix}
    \tilde{\alpha}_{k}\\ \tilde{\beta}_{k}\\ \tilde{\gamma}_{k}
  \end{pmatrix}.
\end{equation}

Since we are interested in $\Box^N\tensor{R}{^\mu_\nu}$, we have to calculate the coefficients $\tilde{\alpha}_0$, $\tilde{\beta}_0$ and $\tilde{\gamma}_0$, which, given Eq.~\eqref{eq:new_recurrence}, turn out to be
\begin{align}\label{eq:tildeCNC0}
    \tilde{\alpha}_0&=(2N)!\,\tilde{\alpha}_N\nonumber\\
    \tilde{\beta}_0&=(2N)!\,\tilde{\beta}_N\\
    \tilde{\gamma}_0&=-\frac{N+1}{2N-1}(2N)!\,\tilde{\gamma}_N.\nonumber
\end{align}
From Eq.~\eqref{eq:odd_term_box_ricci} with $k=0$, $\Box^N\tensor{R}{^\mu_\nu}\,\Box^N\tensor{R}{^\nu_\mu}$ at $r=0$ diverges as
\begin{equation}
  \label{eq:divergence_of_ricci2}
  \Box^N\tensor{R}{^\mu_\nu}\,\Box^N\tensor{R}{^\nu_\mu}=\left(\alpha_0^2+\beta_0^2+2\gamma_0^2\right)r^{-2}+{\cal O}\!\left(r^{-1} \right)\!,
\end{equation}
and combining Eqs.~\eqref{eq:initialCs},~\eqref{eq:relationCs},~\eqref{eq:tildeCNC0} and~\eqref{eq:divergence_of_ricci2} we obtain
\begin{equation}
  \begin{split}
    \Box^N\tensor{R}{^\mu_\nu}\,\Box^N\tensor{R}{^\nu_\mu}= &\ \frac{1}{2}\,(N+1)^2\,(2N)!^2\\
    &\hspace{1.20mm}\boldsymbol{\cdot}\left[11\,b_{2N+1}^2+3(2N+1)\left((2N+1)\frac{a_{2N+1}}{a_0}-2b_{2N+1}\right)\frac{a_{2N+1}}{a_0}\right]r^{-2}\\
    &\hspace{0.35mm}+{\cal O}\!\left(r^{-1} \right)\!.
  \end{split}
\end{equation}
In turn, after imposing the relation between $a_{2N+1}$ and $b_{2N+1}$ in Eq.~\eqref{eq:relation}, we find
\begin{align}
  \Box^N\tensor{R}{^\mu_\nu}\,\Box^N\tensor{R}{^\nu_\mu}=\frac{11}{2}(N+1)^2\,(2N)!^2\,b_{2N+1}^2\,r^{-2}+{\cal O}\!\left(r^{-1} \right)\!.
\end{align}
To conclude, in case at least one of $a_{2N+1}$ and $b_{2N+1}$ is non-vanishing, we always obtain a diverging curvature invariant, which is either $\Box^N{R}$ if Eq.~\eqref{eq:relation} does not hold or $\Box^N\tensor{R}{^\mu_\nu}\,\Box^N\tensor{R}{^\nu_\mu}$ if Eq.~\eqref{eq:relation} holds. This proves our claim.\bigbreak

To recap, in this section we have shown that if any of the conditions of Thm.~\ref{thm:main}~---~$m=n=0$, $b_0=1$, and $\tilde{A}(r)$, $\tilde{B}(r)$ d-even functions~---~is not met, then at least one curvature invariant diverging at $r=0$ exists; thus proving, by contraposition, that the finiteness of all curvature invariants requires these conditions to be satisfied. If either $m\neq0$, $n\neq0$, or $b_0\neq1$, we have seen that at least one of $R$ and $R_{\mu\nu\rho\sigma}R^{\mu\nu\rho \sigma}$ diverges. Instead, if $m=n=0$, $b_0=1$, but $\tilde{A}(r)$, $\tilde{B}(r)$ are not d-even, namely, Eq.~\eqref{eq:expansion_AB} is satisfied, then at least one of the two invariants $\Box^N R$ and $\Box^N \tensor{R}{^\mu_\nu}\Box^N \tensor{R}{^\nu_\mu}$ is divergent. In the case $n_a\neq n_b$, Eq.~\eqref{eq:relation} does not apply and the former invariant is always divergent. Whereas, in the case $n_a=n_b=N$, the relation in Eq.~\eqref{eq:relation} may be satisfied; if it is (not), then the latter (former) invariant diverges.

Note that curvature invariants other than those here examined are expected to diverge as well. For example, if either $m\neq0$, $n\neq0$, or $b_0\neq1$, divergences are found also in higher order contractions of the Ricci tensor, e.g.~$\left(\tensor{R}{^{\mu}_{\nu}}\right)^N$. Similarly, if either $a_{2N+1}\neq0$ or $b_{2N+1}\neq0$, we expect divergences also in other scalars built by contracting $2N$ covariant derivatives with a curvature tensor, such as $\left(\nabla_{\mu_1}\dots\nabla_{\mu_{2N}}R_{\alpha\beta\gamma\delta}\right)^2$. This reasoning is motivated by simple power-counting arguments, through which one can infer whether a curvature invariant constructed with a given number of derivatives of the metric diverges as $r^{-1}$ (or worse). The subtlety lies in the fact that the coefficient multiplying this divergent power might vanish if a certain algebraic relation between $a_{2N+1}$ and $b_{2N+1}$ is encountered (see e.g.~Eq.~\eqref{eq:relation}).

\subsection{A remark on the action of box on rank-two tensors}
We may now address a subtlety that we have omitted in the discussion for the sake of clarity.  On a superficial inspection of Eqs.~\eqref{eq:f1} and~\eqref{eq:f2}, which determine the components of $\Box\tensor{F}{^\mu_\nu}$ from the ones of $\tensor{F}{^\mu_\nu}$, there is no clear reason why the last summand in these expressions should not produce a term proportional to $r^{-2}$ in $\bar{f}_1$ and $\bar{f}_2$, as this would happen in case the terms proportional to $r^0$ in $f_1$ and $f_2$ were different from each other. This would pose a problem in Eq.~\eqref{eq:divergence_of_ricci2}, as this formula would need to incorporate higher order divergent terms that would destroy our efforts to find a relation solely between $a_{2N+1}$ and $b_{2N+1}$. As it turns out, for the tensors we are considering, the terms in $f_1$ and $f_2$ proportional to $r^0$ are actually the same, and we do not get any divergence proportional to $r^{-2}$ in the components of these tensors. Proving this fact explicitly can be quite cumbersome, but we can immediately justify it in light of the results of Sec.~\ref{sec:rev}.

Since we are considering $\tensor{F}{^\mu_\nu}$ to be one of the tensors $\tensor{R}{^\mu_\nu}$, $\Box\tensor{R}{^\mu_\nu}$, \dots~$\Box^{N-1} \tensor{R}{^\mu_\nu}$, all the terms proportional to $r^0$ in $f_1$ and $f_2$, given our starting assumption in Eq.~\eqref{eq:expansion_AB}, arise from combinations of the coefficients proportional to $r^0$, $r^2$, \dots~$r^{2N-2}$ from the Taylor expansion at $r=0$ of $\tensor{R}{^\mu_\nu}$.\footnote{Note that the action of $\Box^{N-1}$ on $\tensor{R}{^{\mu}_{\nu}}$ makes the coefficients of the $r^{2N-1}$ terms to appear at the order $r^1$, which clearly gives no contribution to $r^0$. Thus, the only terms in $\tensor{R}{^{\mu}_{\nu}}$ that, after $\Box^{N-1}$ acts on the Ricci tensor, can give contributions to $r^0$, are terms of lower order than $r^{2N-1}$, i.e. only the even order terms.} We are not interested in what these combinations look like, but it suffices to show that they are solely functions of the even coefficients of $A(r)$ and $B(r)$. Indeed, if we show this, the curvature invariant $\Box\tensor{F}{^\mu_\nu}\Box\tensor{F}{^\nu_\mu}$ would then be divergent if the terms proportional to $r^0$ in $f_1$ and $f_2$ were different. This divergence would contradict the results of Sec.~\ref{sec:rev}, where we have proved that the even order terms in the expansions of $A(r)$ and $B(r)$ cannot produce any diverging curvature invariant, as long as we also assume $m=n=0$ and $b_0=1$.

The fact that the coefficients proportional to $r^0$, $r^2$, \dots~$r^{2N-2}$ in the expansions of the components of $\tensor{R}{^\mu_\nu}$ are only functions of the even coefficients of $A(r)$ and $B(r)$ can be inferred by a direct investigation of Eqs.~\eqref{eq:ricci_components}. There, it is possible to see that the first even order term to which the odd coefficients of $A(r)$ and $B(r)$ contribute is at least $r^{4N}$, which is quadratic in these odd coefficients. Note that, for all $N\geq 0$, $4N>2N-2$, implying that there are no odd coefficients in the terms proportional to $r^0$ in $f_1$ and $f_2$. Therefore, given the results of Sec.~\ref{sec:rev}, for the tensors we are considering, the terms proportional to $r^0$ in $f_1$ must be equal to the ones in $f_2$.

\section{Generalizations of the main theorem}\label{sec:extensions}
The main result of this work, Thm.~\ref{thm:main}, can be reformulated and expanded in terms of $C^k$-extensions of the spacetime, where $k\in\mathbb{N}_0\cup\{\infty\}\cup\{\omega\}$:
\begin{theorem}\label{thm:smooth}
    Given the metric in Eq.~\eqref{eq:metric}, where $A(r)$ and $B(r)$ satisfy the assumptions in Eq.~\eqref{eq:ansatz_metric_functions}, the spacetime is $C^\infty$-extendible to $r=0$ if and only if $m=n=0$, $b_0=1$, and $\tilde{A}(r)$, $\tilde{B}(r)$ are d-even functions of $r$.
\end{theorem}
\begin{theorem}\label{thm:analytic}
    Given the metric in Eq.~\eqref{eq:metric}, where $A(r)$ and $B(r)$ satisfy the assumptions in Eq.~\eqref{eq:ansatz_metric_functions} with $\tilde{A}(r)$ and $\tilde{B}(r)$ analytic functions, the spacetime is $C^\omega$-extendible to $r=0$ if and only if $m=n=0$, $b_0=1$, and $\tilde{A}(r)$, $\tilde{B}(r)$ are even functions of $r$.
\end{theorem}
\begin{theorem}\label{thm:Cn}
    Given the metric in Eq.~\eqref{eq:metric}, where $A(r)$ and $B(r)$ satisfy the assumptions in Eq.~\eqref{eq:ansatz_metric_functions} and are such that $m=n=0$, $b_0=1$, and Eq.~\eqref{eq:expansion_AB} holds (i.e.~at least one of the metric functions has its first non-vanishing odd derivative at order $2N+1$), the spacetime is $C^{2N}$-extendible, but not $C^{2N+2}$-extendible to $r=0$.
\end{theorem}
\noindent Thm.~\ref{thm:smooth} is merely a restatement of Thm.~\ref{thm:main}, and the proof of the former is virtually the same as the proof of the latter. In fact, under the assumptions of Thm.~\ref{thm:main} and in case $m=n=0$, $b_0=1$, and $\tilde{A}(r)$, $\tilde{B}(r)$ d-even at $r=0$, we have already constructed a $C^\infty$-extension of the spacetime to $r=0$ in Sec.~\ref{sec:rev} using the $(t,x,y,z)$ coordinate chart. If one of the above conditions does not hold, we can always construct a diverging curvature invariant that prevents us from obtaining a $C^\infty$-extension to $r=0$.

By the same logic of Thm.~\ref{thm:smooth}, Thm.~\ref{thm:analytic} follows immediately. The only difference stands in checking that the coordinate chart constructed in Sec.~\ref{sec:rev}, under the assumptions of analyticity of $\tilde{A}(r)$ and $\tilde{B}(r)$, produces a $C^{\omega}$-extension of the spacetime to $r=0$. This can be seen from the fact that the metric functions in Eq.~\eqref{eq:extension} are analytic in the $(t,x,y,z)$ coordinates, which, in turn, follows from a straightforward calculation. Note that, in the context of analytic functions, our notions of d-evenness and d-oddness (Def.~\ref{def:d-parity}) coincide with the standard notions of evenness and oddness.

Analogously, also Thm.~\ref{thm:Cn} holds, since, under its assumptions, Sec.~\ref{sec:rev} now provides a Cartesian coordinate chart that allows for continuous derivatives of the metric up to order $2N$, and not $2N+1$, thus realizing a $C^{2N}$-extension of the spacetime to $r=0$. On the other hand, Sec.~\ref{sec:dir} now shows that, in case the Taylor expansions of $A(r)$ and $B(r)$ contain at least one non-vanishing coefficient between $a_{2n_a+1}$ and $b_{2n_b+1}$ where $N\equiv\min\{n_a,n_b\}$, there exist diverging curvature invariants with $2N+2$ derivatives of the metric, thus ruling out the possibility of realizing a $C^{2N+2}$-extension to $r=0$. However, nothing is stated about curvature invariants involving only $2N+1$ derivatives of the metric; hence, whether the spacetime admits a $C^{2N+1}$-extension remains unclear.\footnote{Under the assumptions of Thm.~\ref{thm:Cn}, we empirically observe that all the curvature invariants constructed with $2N+1$ derivatives of the metric, such as $\Box^{N-1}(\nabla^\mu R \nabla_\mu R),\ \Box^{N-1}(\nabla^\rho\tensor{R}{^\mu_\nu} \nabla_\rho\tensor{R}{^\nu_\mu})$ and so on, are finite and continuous at the origin. We turned our attention to the study of other scalar quantities that involve $2N+1$ derivatives of the metric, such as $\Box^{N}\Theta$, $\Box^N \sigma$, and $\Box^N \omega$, where $\Theta$, $\sigma$, and $\omega$ are, respectively, the expansion, shear, and vorticity of a congruence of geodesics~\cite{Hawking:2023lss}. These quantities seem to point in the direction of the $C^{2N+1}$-inextendibility of the spacetime when $a_{2N+1}\neq0$. However, as of now, these results remain inconclusive.}

Note that, as mentioned in Sec.~\ref{sec:intro}, a spacetime can be inextendible past a certain point even if all curvature invariants remain finite at that point. Therefore, asserting that a generic spacetime is extendible to a given point based just on the finiteness of its curvature invariants is, in general, not correct. However, for the static and spherically symmetric geometry here considered, the finiteness of all curvature invariants at $r=0$ is equivalent to $m=n=0$, $b_0=1$, and $\tilde{A}(r)$, $\tilde{B}(r)$ d-even, by Thm.~\ref{thm:main}. In turn, by the results of Sec.~\ref{sec:rev}, the aforementioned conditions on $A(r)$ and $B(r)$ imply the existence of a $C^\infty$-extension to $r=0$~---~or more in general, of a $C^k$-extension, with $k\in\mathbb{N}_0\cup\{\infty\}\cup\{\omega\}$ depending on which of Thms.~\ref{thm:smooth},~\ref{thm:analytic} and~\ref{thm:Cn} is considered. In this context, while the divergence at $r=0$ of a curvature invariant with $k$ derivatives of the metric implies $C^k$-inextendibility, the finiteness of all such invariants is sufficient to guarantee the existence of a $C^k$-extension to $r=0$. In fact, it turns out that the symmetries of a static and spherically symmetric spacetime are strong enough to ensure the validity of also this latter implication. One does not need to check that the components of $R_{\mu\nu\rho\sigma}$ and its covariant derivatives up to order $k-2$ remain finite in a parallelly propagated frame along a curve that approaches $r=0$, as would be needed for a generic spacetime~\cite{Clarke:1973ex,Clarke:1982ex,Clarke:1982ld}. 

If any of the conditions $m=n=0$ and $b_0=1$ does not hold, we know that the spacetime is not $C^2$-extendible to $r=0$~---~since the scalars $R$ and $R_{\mu\nu\rho\sigma}R^{\mu\nu\rho\sigma}$ diverge at that point; however, we cannot use Thm.~\ref{thm:Cn} to determine its degree of extendibility. In this case, we suspect that the spacetime is not even $C^0$-extendible to the origin, although a formal proof would be very technical (see~\cite{Sbierski:2015nta} for the Schwarzschild case). As just mentioned, from the results of Sec.~\ref{sec:dir}, if $m=n=0$ and $b_0\neq1$, $R$ and $R_{\mu\nu\rho\sigma}R^{\mu\nu\rho\sigma}$ diverge at $r=0$ as $r^{-2}$ and $r^{-4}$, respectively. If instead $m=n=0$ and $b_0=1$, but at least one of $a_1$ and $b_1$ is non-vanishing, then Thm.~\ref{thm:Cn} applies and either $R$ or $R_{\mu\nu\rho\sigma}R^{\mu\nu\rho\sigma}$ or both diverge at $r=0$ as $r^{-1}$ and $r^{-2}$, respectively; meaning that the spacetime is $C^0$ but not $C^2$-extendible to the origin.\footnote{Note that, although not shown in Sec.~\ref{sec:dir}, analogous to the case of $\tensor{R}{^\mu_\nu}\tensor{R}{^\nu_\mu}$, $R_{\mu\nu\rho\sigma}R^{\mu\nu\rho\sigma}$ can be easily found to diverge as $r^{-2}$ at $r=0$, with the diverging power multiplied by a coefficient that depends quadratically on $a_1$ and~$b_1$.} In both cases, the singularity at $r=0$ is of the integrable type~\cite{Lukash:2013ts}, as recently mentioned in~\cite{Arrechea:2025fkk}.

\section{Applications and conclusions}\label{sec:conclusions}
The theorems here presented generalize and resolve the conjecture formulated in~\cite{Giacchini:2021pmr} and previously explored in the context of linearized gravity in~\cite{Burzilla:2020utr}. We remark that the results obtained in this work hold in the full non-linear regime, without requiring additional assumptions on the form of the gravitational action. Furthermore, since the metric in Eq.~\eqref{eq:metric} describes very general static and spherically symmetric geometries, with Eq.~\eqref{eq:ansatz_metric_functions} imposing just minimal restrictions on the regularity of $A(r)$ and $B(r)$ at the origin, our results have a wide range of applicability: from (regular) black holes to stellar interiors and, more generally, to any static and spherically symmetric geometry in which $r=0$ is an accessible point. Nevertheless, (regular) black hole spacetimes remain our primary focus, as they are the most likely either to exhibit a singularity at $r=0$ or, in its absence, to fail to admit an extension to $r=0$ with a certain degree of differentiability.

In this context, for a given black hole spacetime, determining whether $m=n=0$, $b_0=1$, and $\tilde{A}(r)$, $\tilde{B}(r)$ are d-even functions is usually a trivial task. Yet, Thm.~\ref{thm:main} and Thm.~\ref{thm:smooth} show that this is equivalent to highly non-trivial statements: namely, the finiteness of all curvature invariants at $r=0$ and, consequently, the existence of a $C^\infty$-extension of the spacetime to the origin. For instance, take the Schwarzschild black hole:
\begin{equation}
	A(r)=\frac{1}{B(r)}=1-\frac{2M}{r},
\end{equation}
for which it is immediate to check that $m=-1\neq0$ and $n=1\neq0$. Then, the results of Sec.~\ref{sec:dir} guarantee the existence of a diverging curvature invariant at $r=0$, which for this geometry is the Kretschmann scalar $R_{\mu\nu\rho\sigma}R^{\mu\nu\rho\sigma}$, diverging as $r^{-6}$ at the origin.

Consider now a black hole spacetime featuring an integrable singularity at $r=0$. Specifically, let us focus on the coherent quantum black hole introduced in~\cite{Casadio:2021eio}, whose metric reads:
\begin{equation}
	A(r)=\frac{1}{B(r)}=1-\frac{2M}{r}\frac{2}{\pi}\text{Si}\hspace{-0.8mm}\left(\frac{r}{R_{\rm s}}\right)\!, \quad R_{\rm s}>0,
\end{equation}
where ${\rm Si}(x)$ denotes the sine integral function. Unlike Schwarzschild, this spacetime is characterized by $m=n=0$, yet
\begin{equation}
	a_0=\frac{1}{b_0}=1-\frac{4M}{\pi R_{\rm s}}\neq1,
\end{equation}
for any finite $R_{\rm s}$. Therefore, from the results of Sec.~\ref{sec:dir} and the above discussion, it follows that curvature invariants such as $R$ and $R_{\mu\nu\rho\sigma}R^{\mu\nu\rho\sigma}$ diverge at $r=0$ as $r^{-2}$ and $r^{-4}$, respectively, indicating a milder singularity at that point.

In the case of regular black hole spacetimes, where typically $m=n=0$ and $b_0=1$ by assumption, Thm.~\ref{thm:Cn} proves particularly useful in determining the degree of differentiability of the spacetime at $r=0$. For example, consider the Hayward black hole~\cite{Hayward:2005gi}:
\begin{equation}
	A(r)=\frac{1}{B(r)}=1-\frac{2Mr^2}{r^3+2ML^2},
\end{equation}
for which $m=n=0$, $a_0=b_0=1$, and the first non-vanishing odd coefficients, $a_{2N+1}$ and $b_{2N+1}$, appear at the order $N=2$: 
\begin{equation}
  a_5=-b_5=\frac{1}{2ML^4}\neq0.
\end{equation}
In light of Thm.~\ref{thm:Cn}, this spacetime can be $C^4$ but not $C^6$-extended to the origin, meaning that every curvature invariant with up to four derivatives of the metric remains finite at $r=0$; this includes e.g.~$R$, $\tensor{R}{_{\mu\nu\rho\sigma}}\tensor{R}{^{\mu\nu\rho\sigma}}$, $\nabla_\alpha \tensor{R}{^{\mu}_{\nu}}\nabla^\alpha\tensor{R}{^{\nu}_{\mu}}$, $\Box R$, and so on. This also means that some curvature invariants with six derivatives of the metric are divergent at $r=0$, e.g.~$\Box^2 R$. These results, which have been already observed in~\cite{Giacchini:2021pmr}, find here a rigorous mathematical foundation.

Instead, in the case of the modified Hayward black hole~\cite{DeLorenzo:2014pta}~---~which incorporates one-loop quantum corrections to the Newtonian potential into the original Hayward spacetime~---~the metric functions read:
\begin{equation}
  A(r)=\left(1-\frac{\alpha\beta M}{\alpha r^3+\beta M}\right)\left(1-\frac{2Mr^2}{r^3+2ML^2}\right)\!,\quad B(r)=\left(1-\frac{2Mr^2}{r^3+2ML^2}\right)^{-1}, \quad\alpha\in(0,1),
\end{equation}
and it is immediate to check that $m=n=0$, $a_0=1-\alpha\neq0$, and $b_0=1$, while for this geometry the first non-vanishing odd coefficient, $a_{2N+1}$, appears at $N=1$:
\begin{equation}
  a_3=\frac{\alpha^2}{\beta M}\neq0, \quad b_3=0.
\end{equation}
This spacetime is less differentiable than the original Hayward spacetime~\cite{Hayward:2005gi}, since it can be $C^2$ but not $C^4$-extended to the origin. In the modified Hayward spacetime, all curvature invariants constructed without additional covariant derivatives are finite at $r=0$, e.g.~$R$ and $\tensor{R}{_{\mu\nu\rho\sigma}}\tensor{R}{^{\mu\nu\rho\sigma}}$; however, already with the inclusion of two covariant derivatives one obtains diverging curvature invariants, e.g.~$\Box R$.

In the previous cases, we have seen only instances of spacetimes that, if extendible to $r=0$ (from a neighborhood of this point) with a certain degree of differentiability, could be extended from both sides: $r>0$ and $r<0$. While being mainly of mathematical rather than physical interest, for suitable $A(r)$ and $B(r)$, it is not a priori forbidden to consider observers approaching the origin also from the ``universe'' where $r<0$, and not just $r>0$.\footnote{We point out that, in our analysis, we are interested only in local extensions of the geometry to $r=0$. For example, an observer from the $r<0$ asymptotic region of the (modified) Hayward spacetime cannot proceed beyond the point at which $r^3=-2ML^2$~\cite{Zhou:2022yio}. Nevertheless, an observer already past this point can still approach $r=0$ from the left.} Nonetheless, in the literature, it is customary to construct extensions to $r=0$ only from the physical $r>0$ portion of the spacetime. An example of this, as mentioned in~\cite{Simpson:2019mud}, is provided by the regular black hole proposed in~\cite{Culetu:2013fsa,Culetu:2014lca}:
\begin{equation}\label{eq:culetu}
    A(r)=\frac{1}{B(r)}=1-\frac{2M}{r} \,\exp\hspace{-0.8mm}\left({-\frac{\alpha}{r}}\right)\!, \quad \alpha>0.
\end{equation}
These metric functions do not satisfy the assumptions in Eq.~\eqref{eq:ansatz_metric_functions}, since~$\lim_{r\to0^-}A(r)=\infty$ faster than any power of $r$. Hence, we cannot directly apply our theorems to infer the degree of extendibility of the spacetime to that point. However, for $A(r)$ and $B(r)$ that satisfy Eq.~\eqref{eq:ansatz_metric_functions} only from the right~---~namely, when Eq.~\eqref{eq:ansatz_metric_functions} holds only in a region $r\in(0,r_0)$, and all derivatives of $\tilde{A}(r)$ and $\tilde{B}(r)$ are finite in the limit $r\to 0^+$~---~we can construct auxiliary metric functions, $\hat{A}(r)$ and $\hat{B}(r)$, that are defined in a neighborhood of $r=0$, such that they coincide with $\tilde{A}(r)$ and $\tilde{B}(r)$ for $r>0$, and such that they are smooth around $r=0$. These auxiliary functions always exist by Borel's lemma, see~\cite{Hormander:2003pdo} Thm.~1.2.6, and provide a smooth continuation of the metric functions to $r\leq0$. The new metric, obtained by replacing $\tilde{A}(r)$ and $\tilde{B}(r)$ with $\hat{A}(r)$ and $\hat{B}(r)$, is equal to the original metric in Eq.~\eqref{eq:metric} only in the region $r>0$, where the coordinate chart $(t,x,y,z)$ from Sec.~\ref{sec:rev} can be constructed independently of the behavior of the metric in the region $r<0$.\footnote{One can check that, by Eq.~\eqref{eq:rtilde} with $b_0=1$, $r>0$ ($r<0$) if and only if $\tilde{r}>0$ ($\tilde{r}<0$). Then, by Eq.~\eqref{eq:xyz}, the coordinates $(t,x,y,z)$ can cover and extend to $r=0$ only one of the two regions $r>0$ or $r<0$ at a time; which region depends on the choice of plus or minus sign in $\tilde{r}=\pm\sqrt{x^2+y^2+z^2}$.} 
This trick enables a straightforward generalization of our theorems to metric functions satisfying Eq.~\eqref{eq:ansatz_metric_functions} only from the right. The main difference is that we are now allowed to discuss the extendibility to $r=0$ solely from the side where $r>0$, since, for $r<0$, the new metric, being merely a mathematical construct, does not need to coincide with the original one. For the metric functions in Eq.~\eqref{eq:culetu}, we find that their extensions are smooth but non-analytic, with $m=n=0$, $a_0=b_0=1$, and $a_k=b_k=0$, $\forall k\geq 1$. Therefore, by Thm.~\ref{thm:smooth}, this spacetime can be $C^\infty$-extended to $r=0$ from the right, which also means that all curvature invariants remain finite in the limit $r\to0^+$.

The above analysis is consistent with the geometrical intuition that the regions $r>0$ and $r<0$ can be extended to $r=0$ independently of one another.\footnote{We want to remark that, if the spacetime can be extended to the origin from both $r>0$ and $r<0$, then we are not joining the two regions through a common $r=0$. A priori, this point should actually be regarded as two distinct points, each belonging to its spacetime region and not the other. The extensions we perform in this work are merely to the point $r=0$ and not beyond.} In fact, assuming $\tilde{A}(r)$ and $\tilde{B}(r)$ to be smooth in a neighborhood of $r=0$ is a mathematical assumption which, in principle, could be relaxed by requiring $\tilde{A}(r)$ and $\tilde{B}(r)$ to be smooth only in one of the two regions, $r>0$ or $r<0$. This applies directly to the case of the last black hole we have examined, where the spacetime is extendible in the only physically viable limit $r\to0^+$. A similar scenario occurs whenever Eq.~\eqref{eq:ansatz_metric_functions} is satisfied with $m,n\in\mathbb{R}\setminus\mathbb{Z}$, i.e.~for non-integer real $m,n$. In this case, the metric functions become complex valued in the region $r<0$. Nevertheless, we can still address the extendibility of the spacetime to $r=0$ from the side $r>0$, where the metric functions remain real valued.\bigbreak

While in general relativity the relevant curvature invariants are the ones that contain two derivatives of the metric, whose finiteness (for a static and spherically symmetric geometry) is guaranteed by a spacetime that is $C^2$-extendible to $r=0$, in the perturbative approach to quantum gravity one might be interested in the finiteness of higher-derivative curvature invariants as well. In this framework, a spacetime that is only $C^2$-extendible is not sufficient to guarantee the finiteness of such invariants at the origin, as recently discussed in~\cite{Koshelev:2024wfk} in the context of Schwarzschild regularizations. In this regard, the theorems presented in this work can find applicability also in the context of higher-derivative theories of gravity, where these higher order curvature invariants are included in the action~\cite{Asorey:1996hz,Modesto:2011kw}.

In particular, Thm.~\ref{thm:Cn} provides an explicit correspondence between the order of the first odd non-vanishing coefficients in the expansions of the metric functions at $r=0$ and the maximum admissible derivative order of the metric in the action before a diverging curvature invariant is produced. In light of the finite action principle~\cite{Barrow:1988act}~---~which states that the only admissible metrics at the quantum level are those that make the action functional in the gravitational path integral finite~---~fixing the maximum derivative order in the action might require imposing a certain d-evenness on the admissible metrics up to a specific order in their Taylor expansions at $r=0$, thereby ensuring the absence of any divergent contribution.\footnote{It is worth noting that, due to the presence of the factor $\sqrt{|g|}$, the action integral may remain finite even if there are divergent curvature invariants~\cite{Giacchini:2021pmr,Borissova:2023kzq}. In fact, for a static and spherically symmetric metric satisfying the assumptions of Eq.~\eqref{eq:ansatz_metric_functions}, the $r$-dependent part of $\sqrt{|g|}$~at $r=0$ goes like $r^{(m+n)/2+2}$. This implies that a curvature invariant with a sufficiently mild divergence at the origin can still produce a finite contribution to the action once integrated. For example, for $m=n=0$, curvature invariants diverging as $r^{-1}$ or $r^{-2}$ can be integrated in a neighborhood of $r=0$ without making the action to diverge.}

In the specific case of higher-derivative theories of gravity whose actions include curvature invariants of arbitrarily high order~\cite{Modesto:2011kw}, Thms.~\ref{thm:smooth} and~\ref{thm:analytic} may instead prove more useful. In this setting, the only metrics that can regularize curvature invariants containing derivatives of the metric of arbitrarily high order are the ones that are d-even.

\section*{Acknowledgments}
The authors thank two anonymous referees for their helpful comments, which have led to this improved version of the manuscript.

The work of T.A. is supported by a doctoral studentship of the Science and Technology Facilities Council (training grant no.~ST/Y509620/1, project reference no.~2917813).
The work of M.S. is supported by a doctoral studentship of the Science and Technology Facilities Council (training grant no.~ST/X508822/1, project reference no.~2753640).

\appendix
\section{Properties of d-even and d-odd functions}\label{sec:properties}
We here provide the proof of Prop.~\ref{prop:d-parity}. The claims {\it e)} and {\it f)} of the proposition are immediate from Def.~\ref{def:d-parity}. To prove the claims {\it a)}, {\it b)}, {\it c)}, and {\it d)} we introduce the operator $T_a$, which takes a function $f(x)$ that is smooth in a neighborhood of $x=a$ and yields its Taylor series around this point:
\begin{equation}
  T_a f(x)=\sum_{n=0}^\infty \frac{f^{(n)}(a)}{n!}(x-a)^n.
\end{equation}
If the function $f(x)$ is only smooth, then we do not necessarily have $T_a f(x)=f(x)$, which would be true for an analytic $f(x)$. Instead, $T_a f(x)$ must be interpreted simply as a generating function that collects all the derivatives of $f(x)$ at $x=a$. It is not difficult to show that this power series behaves naturally under function composition with another smooth function $g(x)$:
\begin{equation}
  T_a (f\circ g)(x)=T_{g(a)}f(\,T_a g(x)\,),
\end{equation}
as can be verified via a direct calculation using Faà di Bruno's formula~\cite{Johnson:2002fdb}, which expresses the $n$th derivative of the composition of two functions as
\begin{equation}
  \frac{d^n}{dx^n} (f\circ g)(x)=\mathlarger{\sum}_{\substack{m_1,\,\dots\,m_{n}\geq0\\1\cdot m_1+\,\dots\,+n\cdot m_{n}=n}} \frac{n!}{m_1!\,\cdots\, m_n!}\,f^{(m_1+\,\dots\,+m_n)}(g(x))\,\prod_{k=1}^n\left(\frac{g^{(k)}(x)}{k!}\right)^{m_k}.
\end{equation}
Indeed, we obtain that
\begin{equation}
\begin{split}
  &T_{g(a)}f(\,T_a g(x)\,)=\\
  &=\sum_{n=0}^\infty \frac{f^{(n)}(g(a))}{n!}\left(\sum_{m=0}^\infty \frac{g^{(m)}(a)}{m!}(x-a)^m-g(a)\right)^n\\
  &=\sum_{n=0}^\infty \frac{f^{(n)}(g(a))}{n!}\left(\sum_{m=1}^\infty \frac{g^{(m)}(a)}{m!}(x-a)^m\right)^n\\
  &=\sum_{n=0}^\infty \frac{f^{(n)}(g(a))}{n!}\left[\mathlarger{\sum}_{\substack{m_1,m_2\,\dots\,\geq0\\m_1+m_2+\,\dots\,=n}}\frac{n!}{m_1!\,m_2!\,\cdots} \prod_{k=1}^\infty\left(\frac{g^{(k)}(a)}{k!}(x-a)^k\right)^{m_k}\right]\\
  &=\sum_{n=0}^\infty \left[\mathlarger{\sum}_{\substack{m_1,m_2\,\dots\,\geq0\\m_1+m_2+\,\dots\,=n}}\frac{f^{(m_1+m_2+\,\dots)}(g(a))}{m_1!\,m_2!\,\cdots} \prod_{k=1}^\infty\left(\frac{g^{(k)}(a)}{k!}\right)^{m_k}(x-a)^{1\cdot m_1+2\cdot m_2+\,\dots}\right]\\
  &=\sum_{n'=0}^\infty\frac{1}{n'!}\hspace{-0.125cm}\left[\mathlarger{\sum}_{\substack{m_1,\,\dots\,m_{n'}\geq0\\1\cdot m_1+\,\dots\,+n'\cdot m_{n'}=n'}} \hspace{-0.1cm}\frac{n'!}{m_1!\,\cdots\,m_{n'}!} f^{(m_1+\,\dots\,+m_{n'})}(g(a)) \prod_{k=1}^{n'}\left(\frac{g^{(k)}(a)}{k!}\right)^{m_k}\right]\hspace{-0.125cm}(x-a)^{n'}\\
  &=\sum_{n'=0}^\infty \frac{(f\circ g)^{(n')}(a)}{n'!}(x-a)^{n'}\\
  &=T_{a}(f\circ g)(x),
\end{split}
\end{equation}
where in this proof we have just used the multinomial theorem, see e.g.~\cite{Stanley:2011ecv} Sec.~1.2, and reindexed the main sum from $n$ to $n'$. In the above, $n$ represents the order of the derivative of $f(x)$, while $n'$ corresponds to the order of the power of the $(x-a)$ factor. The latter index is therefore more suited in order to resum the expression as a Taylor series. Moreover, the reindexing has the effect of turning the condition on the internal sum from $m_1+m_2+\,\dots\,=n$ to $1\cdot m_1+\,\dots\,+n'\cdot m_{n'}=n'$, effectively reducing the number of indices $m_k$ from an infinite amount to only $n'$ of them.\newpage

To prove {\it a) $f$ generic, $g$ d-even $\implies f\circ g$ d-even}, we have a d-even function $g$ and a generic smooth function $f$, for which
\begin{align}
  T_0 g(x)&\equiv\sum_{n=0}^\infty g_{2n}\, x^{2n},\\
  T_{g_{0}}f(x)&\equiv\sum_{n=0}^\infty f_n (x-g_{0})^n.
\end{align}
We can calculate the structure of the derivatives of $f\circ g$ via 
\begin{equation}
\begin{split}
  T_0 (f\circ g)(x)=T_{g_{0}}f(\,T_0 g(x)\,)&=\sum_{n=0}^\infty f_n \left(\sum_{m=0}^\infty g_{2m}\, x^{2m}-g_{0}\right)^n\\
  &=\sum_{n=0}^\infty f_n \left(\sum_{m=1}^\infty g_{2m}\, x^{2m}\right)^n\\
  &\equiv\sum_{k=0}^\infty h_{2k}\, x^{2k}.
\end{split}
\end{equation}
Even though an explicit expression for $h_{2k}$ is cumbersome, it is clear that the power series is even in $x$, which means that $f\circ g$ is d-even.\bigbreak

To prove {\it b) $f$ d-even, $g$ d-odd $\implies f\circ g$ d-even}, we have a d-odd function $g$ and a d-even function $f$, for which
\begin{align}
  &T_0 g(x)\equiv\sum_{n=0}^\infty g_{2n+1}\, x^{2n+1},\\
  &T_{0}f(x)\equiv\sum_{n=0}^\infty f_{2n}\, x^{2n}.
\end{align}
Again, the structure of the derivatives of $f\circ g$ is given by
\begin{equation}
\begin{split}
  T_0 (f\circ g)(x)=T_{0}f(\,T_0 g(x)\,)&=\sum_{n=0}^\infty f_{2n} \left(\sum_{m=0}^\infty g_{2m+1}\, x^{2m+1}\right)^{2n}\\
  &=\sum_{n=0}^\infty f_{2n} \,x^{2n}\left(\sum_{m=0}^\infty g_{2m+1}\, x^{2m}\right)^{2n}\\
  &\equiv\sum_{k=0}^\infty h_{2k}\, x^{2k}.
\end{split}
\end{equation}
And yet again, clearly $f\circ g$ is d-even. To prove {\it c) $f$ d-odd, $g$ d-odd $\implies f\circ g$ d-odd}, we follow the same logic, so we omit the proof here.\bigbreak

To prove {\it d) $f$ d-odd $\iff f^{-1}$ d-odd}, we know that $f$ is d-odd, hence 
\begin{equation}
  T_{0}f(x)\equiv\sum_{n=0}^\infty f_{2n+1}\, x^{2n+1},
\end{equation}
and moreover we assume $f_{1}\neq 0$, so that by the inverse function theorem $f$ has a smooth inverse, see e.g.~\cite{Tao:2022ana} Sec.~6.7 and~\cite{Hormander:2003pdo} Thm.~1.1.7. We want to prove that $f^{-1}$ is also d-odd; therefore, by way of contradiction, we suppose that there exists an $\ell$ such that $f^{-1\,(2\ell)}(0)\neq0$, and let us take this~$\ell$ to be minimal, i.e.~$f^{-1\,(2k)}(0)=0$ for all $k$ strictly smaller than $\ell$. We can thus write: 
\begin{equation}
  T_{0}f^{-1}(x)\equiv\sum_{n=0}^{\ell-1} \tilde{f}_{2n+1}\, x^{2n+1}+\sum_{n=2\ell}^\infty\tilde{f}_{n}\, x^{n}, \qquad \tilde{f}_{2\ell}\neq 0, 
\end{equation}
but we also know that $f^{-1}\circ f=\text{id}$, with $\text{id}$ the identity function. If we use the rule for compositions, we get
\begin{equation}
\begin{split}
  x&=T_0\text{id}(x)=T_0(f^{-1}\circ f)(x)=T_{0}f^{-1}(\,T_{0}f(x)\,)\\
  &=\sum_{n=0}^{\ell-1} \tilde{f}_{2n+1}\, \left(\sum_{m=0}^\infty f_{2m+1}\, x^{2m+1}\right)^{2n+1}+\sum_{n=2\ell}^\infty\tilde{f}_{n}\, \left(\sum_{m=0}^\infty f_{2m+1}\, x^{2m+1}\right)^{n}\\
  &=\sum_{n=0}^{\ell-1} \tilde{f}_{2n+1}\, x^{2n+1}\left(\sum_{m=0}^\infty f_{2m+1}\, x^{2m}\right)^{2n+1}+\sum_{n=2\ell}^\infty\tilde{f}_{n}\, x^n\left(\sum_{m=0}^\infty f_{2m+1}\, x^{2m}\right)^{n}\\
  &\equiv \sum_{n=0}^{\ell-1} h_{2n+1}\, x^{2n+1}+\tilde{f}_{2\ell}\,(f_1)^{2\ell}\, x^{2\ell}+\sum_{n=2\ell+1}^\infty h_{n}\, x^{n}.
\end{split}
\end{equation}
In practice, in the rhs we obtain even powers of $x$, the smallest of which is $x^{2\ell}$ with a non-vanishing coefficient $\tilde{f}_{2\ell}\,(f_1)^{2\ell}$, but this is in contradiction with the lhs where we have only a contribution of~$x^1$. We have thus proved that $f^{-1}$ is d-odd. The converse implication is true by symmetry.

\printbibliography
\end{document}